\documentclass[10pt,twocolumn]{article}
\usepackage[utf8]{inputenc}
\usepackage[english]{babel}
\usepackage[numbers]{natbib}
\usepackage[T1]{fontenc}
\usepackage[top=2cm, bottom=2cm, left=1.5cm,right=1.5cm]{geometry}

\usepackage{graphicx}%
\usepackage{xurl}
\usepackage[ruled]{algorithm2e}
\usepackage{algpseudocode}
\graphicspath{{./figures/}}
\usepackage{multirow}%
\usepackage{amsmath,amssymb,amsfonts}%
\usepackage{amsthm}%
\usepackage{mathrsfs}%
\usepackage[title]{appendix}%
\usepackage{xcolor}%
\usepackage{textcomp}%
\usepackage{manyfoot}%
\usepackage{booktabs}%
\usepackage{listings}%
\usepackage{tikz,lipsum,lmodern}
\usepackage{amsmath,amssymb,amsthm}

\theoremstyle{definition}

\usepackage{enumerate}
\usepackage{qcircuit}

\usepackage{caption}
\usepackage{subcaption}
\usepackage[most]{tcolorbox}

\usepackage{authblk}

\newcommand{\ket}[1]{\ensuremath{\left|#1\right\rangle}} 
\newcommand{\bra}[1]{\ensuremath{\left\langle#1\right|}} 
\newcommand{\braket}[2]{\ensuremath{\left\langle#1|#2\right\rangle}}

\newcommand{\revfix}[1]{{#1}}
\title{Error analysis of quantum operators written as a linear combination of permutations}
\author{Ammar Daskin\thanks{adaskin25@gmail.com}}
\affil{Dept. of Computer Engineering,\\
Istanbul Medeniyet University,\\ Istanbul, Turkiye, 34000.
}

\date{}
\begin{document}
\maketitle

\begin{abstract}
In this paper, we consider matrices given as a linear combination of permutations and analyze the impact of bit and phase-flips on the perturbation of the eigenvalues.
When the coefficients in the linear combination are positive, we observe that the eigenvalues of the resulting matrices exhibit resilience to quantum bit-flip errors.
In addition, we analyze the bit-flips in combination with positive and negative coefficients and the phase-flips.  Although matrices with mixed-sign coefficients show less resilience to the bit-flip and phase-flip errors, the numerical evidence shows that the perturbation of the eigenspectrum is very small when the rate of these errors is small.
We also discuss the situation when this matrix is implemented through block encoding and there is a control register.
Since any square matrix can be expressed as a linear combination of permutations multiplied by two scaling matrices from the left and right (via Sinkhorn’s theorem), this paper gives a framework to study matrix computations in quantum algorithms related to numerical linear algebra.
In addition, it can give ideas to design more error-resilient algorithms that may involve quantum registers with different error characteristics.
\end{abstract}

\section{Introduction}
 In quantum mechanics and many other fields, perturbation theory is used to find approximate solutions for systems or to study system dynamics under perturbations \cite{kato2013perturbation,greenbaum2020first}.
Such analysis can be done by considering perturbations on the input data or on the system.
For a given matrix $A\in C^N$ with $A\ket{\psi} = \lambda \ket{\psi}$ where $\lambda$ is a simple eigenvalue and \ket{\psi} is the corresponding eigenvector; 
the eigenvalue perturbation analysis \cite{kato2013perturbation,greenbaum2020first} is the study of understanding how $\lambda$ changes when the small changes occur in the matrix elements of $A$ \cite{golub2000eigenvalue}.
In general, bounding the error in eigenvalue perturbation in terms of changes in matrix elements is difficult because even changing a single element (e.g., setting one element to zero) can drastically alter eigenvalues (see Theorem III.2.1 in \cite{bhatia2007perturbation}; see also  Chapter 8 in \cite{golub2013matrix} for related discussion).

Therefore, the Gershgorin circle theorem, which provides bounds on the locations of the eigenvalues, is very useful in practice \cite{golub2000eigenvalue,golub2013matrix}: For a given $n \times n$ matrix $A$, the $i$-th Gershgorin circle is defined as the set of all complex numbers $z$ such that:
$|z - a_{ii}| \leq \sum_{j=1, j\neq i}^{n} |a_{ij}|$, where $a_{ii}$ is the $i$-th diagonal element of $A$ and $a_{ij}$ are the off-diagonal elements. 
Each Gershgorin circle is centered at the diagonal element $a_{ii}$ with a radius equal to the sum of the absolute values of the off-diagonal elements in the same row. 
The theorem implies that the eigenvalues of $A$ are located within the union of these circles.

In classical computation, changes in the matrix elements are caused by natural system limitations such as floating-point representations \cite{wilkinson1960error,veselic1993floating}. 
These errors sometimes accumulate depending on the way the algorithm is designed for the problem or how the algorithm is implemented on computers. As a result, studies on eigenvalue perturbation play a significant role in determining whether, in some cases, these errors can be prevented or minimized by redesigning algorithms or by rewriting implementations in less error-prone ways. For instance, in the Gram-Schmidt process \cite{golub2013matrix}, orthonormalization is achieved by subtracting from each vector its projection onto the directions of previously chosen vectors.
In the classical Gram-Schmidt algorithm, the vectors are orthonormalized one at a time. The reorthogonalization step in the modified version of the algorithm improves the numerical stability of the algorithm and reduces the numerical errors caused by the finite-precision arithmetic.

Since perturbation theory is a frequently applied method in quantum mechanics, similar eigenvalue perturbation analyses are also valid for quantum computation.
 Quantum computation relies on manipulating probability amplitudes, where measurement outcomes depend on squared magnitudes. 
For example, if we measure the right answer on a qubit with probability $p-\epsilon$ instead of $p$, 
it may not affect the algorithmic result when the error $\epsilon$ is not big enough to change the measurement outcome. In other words, small errors in amplitudes do not affect the algorithmic outcome generally defined in terms of classical bits (i.e. 0, 1) when they are below a threshold that alters the most probable result.

While quantum computers inherit the sensitivity to the changes in the matrix elements, as long as these changes do not disrupt the measurement outcome (the algorithmic accuracy) through bit-flips (Pauli-$X$ which flips the state of a qubit) and phase-flips (Pauli-$Z$ which flips the sign of the amplitude), they can be neglected.
This is one reason why error-resilient quantum computational protocols are possible: Examples of such protocols and algorithms include \cite{turkeshi2024error} which shows error resilience phase transitions in quantum circuits, \cite{huo2023error} which presents an error resilient quantum Monte Carlo simulation, and \cite{braun2022error} which gives an error resilient version of the phase estimation algorithm.

On the other hand, while fault-tolerant quantum computation provides theoretical safeguards against these errors, practical implementations of quantum algorithms—particularly in numerical linear algebra—require granular analyses of operator resilience to guide error-mitigation strategies \cite{xiong2021quantum,lowe2021unified} (see \cite{cai2023quantum} for a review of error-mitigation techniques and their limits \cite{takagi2022fundamental}).
\subsection{Main motivation points}
\begin{itemize}
    \item Quantum circuits are the evolution operators of some Hamiltonian operators. Therefore, while the effect of bit-flip and phase-flip errors on the eigenvalues of the Hamiltonian can be studied using random quantum circuits and applying random errors, the model may include errors from mapping the Hamiltonian to a quantum circuit or from finding the evolution operator of the Hamiltonian (e.g., finding circuits through Trotter-Suzuki decomposition).
    \item 
Quantum algorithms related to numerical linear algebra mostly use the block encoding \cite{gilyen2019quantum} scheme to represent a quantum operator since it also allows  for the embedding of non-unitary matrices. 
In block encoding schemes, the operator $A$, which may not be a Hermitian matrix, is generally considered as a linear combination of unitary matrices.
It is then used as part of a larger quantum circuit represented by the following unitary matrix:
\begin{equation}
\label{Eq:blockencoding}
    \mathcal{U} = \left(\begin{matrix}
        A&\bullet\\
        \bullet & \bullet\\
    \end{matrix}\right).
\end{equation}
Here, it is important to represent $A$ in simpler terms instead of general unitaries in order to make the stability analyses of numerical algorithms easier.

\item 
Any bistochastic matrix can be written as a convex combination of permutations (Birkhoff's theorem \cite{birkhoff1946tres}). 
It is also possible to write any matrix $A$ as a linear combination of permutations multiplied by two scaling matrices from the left and right in accordance with Sinkhorn's theorem \cite{sinkhorn1964relationship}: $A = D_1\left( \sum_{i=1}^K \alpha_i \Pi_i \right)D_2$, where $D_1$ and $D_2$ are diagonal scaling matrices, $\Pi_i$ are permutations, $\alpha_i$ are scalar, and $K$ is the number of terms. 
 This result is closely related to the singular value decomposition (SVD) where the scaling matrices are found and the Birkhoff–von Neumann decomposition for bistochastic matrices (e.g., see \cite{kulkarni2017minimum,daskin2024quantum} see also \cite{idel2015sinkhorn} for unitary matrices).
As an example, consider a $2 \times 2$ matrix $A$ which can be written as:
\begin{equation}
    A = D_1 (\alpha_1 \Pi_1 + \alpha_2 \Pi_2) D_2,
\end{equation}
where:
\begin{equation}
      \Pi_1 = \begin{bmatrix} 1 & 0 \\ 0 & 1 \end{bmatrix}, \quad \Pi_2 = \begin{bmatrix} 0 & 1 \\ 1 & 0 \end{bmatrix}.
\end{equation}
By choosing appropriate scaling matrices $D_1$ and $D_2$ and coefficients, we can represent the matrix $A$.
One can also write matrix elements as a sum of Kronecker products of $X$ and controlled-$X$ gates by dividing the matrix into blocks, writing each block as a sum of matrices multiplied by some controlled-$X$ operations, and then repeating the same operations for the terms in the summation \cite{daskin2018generalized}. Note that this does not necessarily generate an efficient decomposition.
\end{itemize} 

\subsection{Main contributions}
In this paper, we consider $A$ as a linear combination of permutations which simplifies error modeling. Since any matrix $A$ can be written as a linear combination of permutations (after a left and a right scaling), this  also retains broad applicability of the block encoding scheme.
Then, we present a probabilistic model of the bit and phase-flips on the permutation terms of the linear combination. In this probabilistic model, we analyze the bound of the eigenvalue perturbation and run numerical simulations to observe the change of the eigenvalue spectrum under the influence of bit and phase-flip errors that occur with different probabilities. 
In addition, we show how the perturbations in the matrix and eigenvalues in this model relate to the perturbations in the fidelity of an input quantum state.

In the analyses of the model, we analyze positive and mixed-sign coefficients separately:
When the linear combination is constructed with positive coefficients, the resulting matrix becomes positive (the matrix elements are non-negative). 
From the Perron-Frobenius theorem, it is known that the non-negative matrices have dominant eigenvalue pairs. 
We show that the matrices written as a linear combination of permutations with positive coefficients are particularly resilient to bit-flip errors.

In the following section, we first describe the model and show the mathematical analysis and numerical simulations for the eigenvalue perturbation.
Then, in Sec.\ref{Sec:fid}, we analyze how the fidelity of a quantum state is affected by the perturbation in the operator in this defined model. And in the final sections we conclude the paper with the discussion of how the presented model can be used to design error resilient quantum algorithms.

\section{The model}
In this paper, we assume that we are given $A$ as a linear combination of $K$ permutations:
\begin{equation}
\label{Eq:A}
    A = \sum_i^K \alpha_i\Pi_i,
\end{equation} 
where $\Pi_i$ is a permutation and $\alpha_i$s  are  complex valued coefficients, $\alpha_i \in C$.
Note that if the coefficients are  real and positive, the matrix becomes positive and it has a dominant eigenvalue.

\subsection{Modeling bit and phase-flip errors in matrix $A$}
The quantum errors \cite{kitaev1997quantum,devitt2013quantum} can be categorized as bit-flip and phase-flip errors. In the case of a permutation $\Pi_i$,  a bit-flip error can result in a different permutation decomposition. We can directly consider a bit-flip error as applying a NOT gate to a qubit before or after the permutation operation. Permutations admit representations like cycle decompositions, permutation matrices, or gate sequences (e.g., SWAPs). Therefore, in addition to applying a NOT gate, we can simulate the bit-flip error by either slightly changing the permutation matrix, permutation string, or cycles if the permutation is written in the form of cycles.

With the probability $p_i$, we have a bit-flip on any random qubit $b_i$ (or any random group of qubits) after applying the permutation $\Pi_i$.
That means we have the term $\alpha_i\Pi_i$ with probability $(1-p_i)$ or $\alpha_iX^{b_i}\Pi_i $ with probability $p_i$: 
Note that $X^{b_i}$ represents a NOT gate on qubit $b_i$ or a group of qubits represented by $b_i$.
Therefore, for a given probability vector $p$ and a vector of qubit-indices $b$, we can represent the applied operator by the following probabilistic model:
\begin{equation}
\begin{split}
\label{Eq:Ap}
    A(p,b) = & \sum_i^K (1-p_i) \alpha_i\Pi_i + p_i \alpha_iX^{b_i}\Pi_i, \\
     = & \sum_i^K \alpha_i\left(\left(1-p_i\right)I+ p_i X^{b_i}\right)\Pi_i.
     \end{split}
\end{equation}

In similar fashion, for each term if we assume a phase-flip on any random qubit $\phi_i$ with probability $q_i$, we can write the following:
\begin{equation}
\label{Eq:Aq}
    A(q,\phi) = \sum_i^K \alpha_i\left(\left(1-q_i\right)I+ q_i Z^{\phi_i}\right)\Pi_i,
\end{equation}
where $q$ and the $\phi$ are the vectors indicating probabilities and the random qubits.  Here, $Z$ denotes the Pauli-$Z$ gate, which applies a phase-flip ($\ket{0} \rightarrow \ket{0}$, $\ket{1} \rightarrow -\ket{1}$).

If there are both phase and bit-flip errors, we can combine the above equations and rewrite the model as:
\begin{equation}
\label{Eq:Apq}
\begin{split}
        A(p,q,b,\phi) =  \sum_i^K &
        \alpha_i\left(\left(1-q_i\right)I+ q_i Z^{\phi_i}\right)\\
        &\left(\left(1-p_i\right)I+ p_i X^{b_i}\right)\Pi_i.
\end{split}
\end{equation}
Here,  following standard quantum error models \cite{nielsen2010quantum}, we assume bit-flips and phase-flips occur independently, with phase-flips applied after bit-flips to model sequential error propagation.

\subsection{Analysis of bit-flips}
\label{Sec:ABitFlips}
\subsubsection{The case $\alpha_i\in R$ and $0\leq \alpha_i \leq 1$}
{For $\alpha_i > 0$, $A$ is non-negative, ensuring a dominant eigenvalue by the Perron-Frobenius theorem \cite{golub2013matrix}. 
Therefore, from the Perron–Frobenius theorem, $A$ has a non-zero eigenvector (the eigenvector elements are real and non-negative) and a positive dominant eigenvalue in the sense that there is a gap between this and the second largest eigenvalue.}
Assume these pairs are the eigenvalue $\lambda_{max}$ and the eigenvector \ket{x} and the dimension of the matrix is $N$: We can write the eigenvalue equation in terms of matrix and vector elements as:
\begin{equation}
\begin{split}
        A\ket{x} & = \lambda_{max} \ket{x}\\ 
    & = \left(\begin{matrix}
    \sum_j^N a_{1j}x_j\\
    \vdots\\
    \sum_j^N a_{Nj}x_j\\
    \end{matrix}\right) = 
\lambda_{max}\left(\begin{matrix}
    x_1\\
    \vdots\\
    x_N\\
\end{matrix}\right). 
\end{split}
\end{equation}

If we sum the elements in both side of the equation, then we obtain:
\begin{equation}
    \sum_{i=1}^N \sum_{j=1}^N a_{ij}x_j = \lambda_{max} \sum_{i=1}^N x_i.
\end{equation}
Since $A$ is a linear combination of permutation matrices scaled by $\alpha_i$, each column sum $s_j = \sum_i a_{ij} = \sum_i \alpha_i$ (as permutations preserve column sums). Thus,  
\begin{equation}
    \sum_{j=1}^N s_j x_j = \lambda_{max} \sum_{j=1}^N x_j.
\end{equation}
For a bistochastic matrix ($s_j = \sum_i \alpha_i$ constant), $\lambda_{max} = \sum_i \alpha_i$, invariant under bit-flips that preserve column sums (permutations preserve column sums for bistochastic matrices). 
By the Perron-Frobenius theorem \cite{golub2013matrix}, the dominant eigenvalue $\lambda_{max}$ of a non-negative matrix is isolated and corresponds to a positive eigenvector. Bit-flips that preserve non-negativity (e.g., permuting entries without sign changes) maintain $\lambda_{max}$ stability.  
Since the dominant eigenvalue is not affected by the bit-flip errors, the applications (e.g., \cite{brualdi1988some}) that are based on this type of matrix are resilient to quantum bit-flip errors.

{ For the other eigenvalues, the same cannot be said since the eigenvectors may have non-positive entries. Since non-dominant eigenvalues correspond to eigenvectors with mixed-sign entries, their sensitivity to perturbations increases. However, when bit-flips occur in terms with small coefficients ($|\alpha_i| \ll \lambda_{max}$), their impact on the spectrum is attenuated. }

We can also analyze this via the Gershgorin circle theorem \cite{golub2013matrix,horn2012matrix}, which indicates that every eigenvalue is bounded inside one of the circles that are defined by the center $a_{ii}$, the $i$th diagonal element of matrix $A$, and the following radius:
\begin{equation}
    R_i = \sum_{j, j\neq i}^{N}|a_{ij}|.
\end{equation}
When bit-flips are applied; for each term, the Gershgorin radius changes by \( \Delta R_i \leq 2|\alpha_i| \), independent of the number of flipped qubits. Summing over all terms, \( \Delta R_j \leq 2\sum_{i=1}^K p_i|\alpha_i| \) (see Appendix \ref{appendixBitFlip} for a detailed derivation). For small $K$ (i.e. $K\ll N$) and $|\alpha_i|$ or for small $p_i$ (i.e. $p_i \ll 1/K$; $\Delta R_i$ remains negligible, leaving eigenvalue bounds largely unchanged.  However, while the  absolute changes might be small, relative changes $\Delta R_i/|\lambda_i|$ could be significant: For eigenvalues with small magnitudes, the relative perturbation 
 may become significant even if $\Delta R_i$ is small.  

\begin{figure}[t!]
    \centering
    \begin{subfigure}[t]{1\columnwidth}
        \centering
        \includegraphics[width=1\columnwidth]{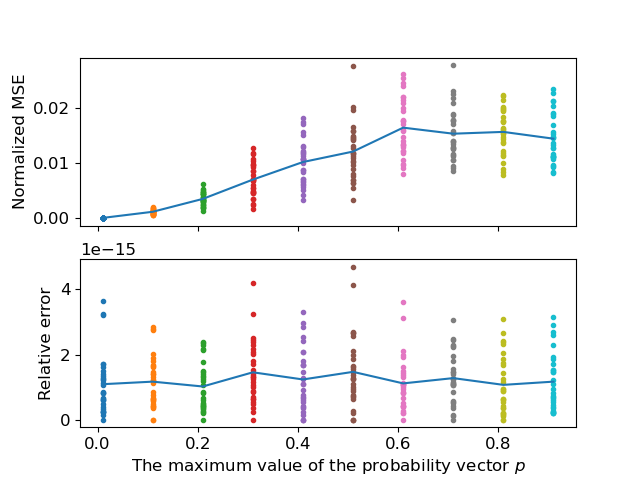}
        \caption{$K = 16$: For each term, $1\leq \#flips \leq 8$ and $0 \leq \alpha_i\leq 1$.}
    \end{subfigure}
    \begin{subfigure}[t]{1\columnwidth}
        \centering
        \includegraphics[width=1\columnwidth]{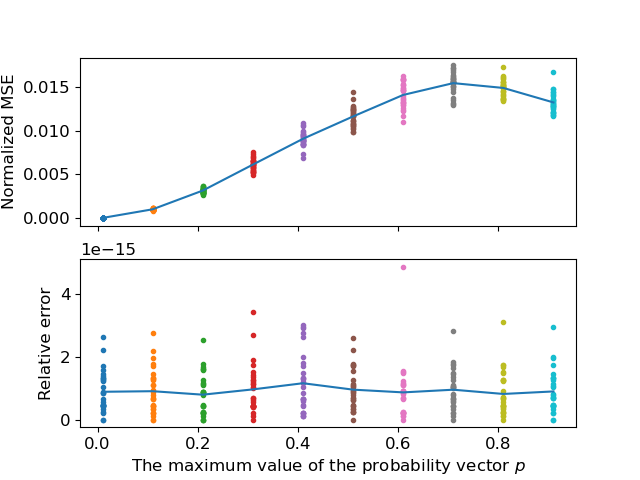}
        \caption{$K = 256$: For each term, $1\leq \#flips \leq 8$ and $0 \leq \alpha_i\leq 1$. }
    \end{subfigure}
       \caption{\label{Fig:bitflips01}
 Relative error (Eq.~\eqref{Eq:relative error}) and NMSE (Eq.~\eqref{Eq:nmse error}) for $A(p,b)$ with $n=8$ qubits: Each trial (dot) applies bit-flips to 1–8 random qubits per term. Lines show mean relative error in Eq.\eqref{Eq:relative error} and NMSE in Eq.\eqref{Eq:nmse error} across 30 trials run with different probability choices in $A(p, b)$. (a) $K=16$; (b) $K=256$.}
    \end{figure}

The numerical simulation of the discussed bit-flip case is presented in Fig.\ref{Fig:bitflips01} where we use random matrices of $n=8$ qubits with the number of terms $K=256$ (i.e. $K = 2^n$)  and, in a separate simulation, $K=16$ (corresponding to $2n$). 
In the figures,  for each term in the summation, we randomly choose probabilities $p_i$ and $q_i$ from the probability range [0, $P_{max}$], and randomly choose group of qubit indices $b_i$ and $\phi_i$ from [1, $G_{max}$]. 
Here, while $P_{max}$ indicates maximum probability, $G_{max}$ indicates the maximum possible vector size of $b_i$ and $\phi_i$, i.e. the maximum number of qubits that are flipped in each term. In these cases, it is set to $n$. 

The changes measured in terms of the relative error in the dominant eigenvalues, $\lambda_{max}(A)$ and e.g. $\lambda_{max}(A(p,q))$, of  the original and the perturbed matrices:
\begin{equation}
\label{Eq:relative error}
    re = \frac{|\lambda_{max}(A)-\lambda_{max}(A(p,q))|}{|\lambda_{max}(A)|},
\end{equation}
and the normalized mean squared error of all eigenvalues:
\begin{equation}
\label{Eq:nmse error}
    NMSE =  \frac{1}{N}\sum_i^N \frac{|\lambda_i(A)-\lambda_i(A(p,q))|^2}{|\lambda_{max}(A)|}.
\end{equation}
    
\subsubsection{The case where we have a control and a system registers}
The implementation of the circuit $A$ via $\mathcal{U}$ given in Eq.\eqref{Eq:blockencoding} generally involves a control register on the permutations.
The control register may also implement the coefficients. If there is a bit flip on the control register, the coefficients of the permutations are swapped, but the eigenvalue spectrum remains the same. Therefore, $A$ built with positive coefficients is resilient to bit-flip errors in both registers.

\subsubsection{The case $\alpha_i\in R$ and $-1\leq \alpha_i \leq 1$}
\begin{figure}[t!]
    \centering
    \begin{subfigure}[t]{1\columnwidth}
        \centering
        \includegraphics[width=1\columnwidth]{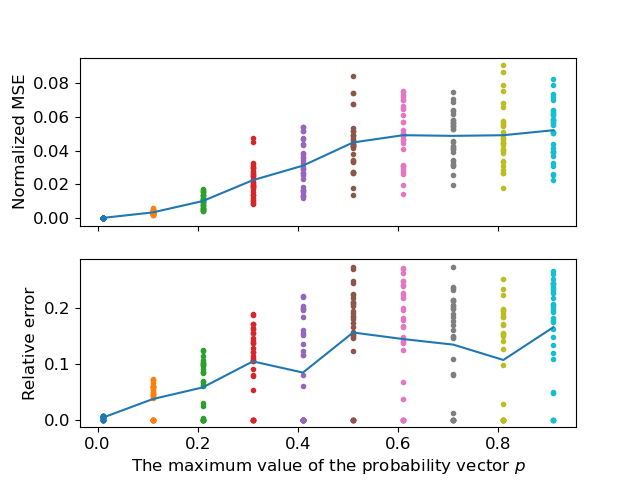}
        \caption{$K=16$: For each term, $1\leq \#flips \leq 8 $ and $-1 \leq \alpha_i\leq 1$.}
    \end{subfigure}
    \begin{subfigure}[t]{1\columnwidth}
        \centering
        \includegraphics[width=1\columnwidth]{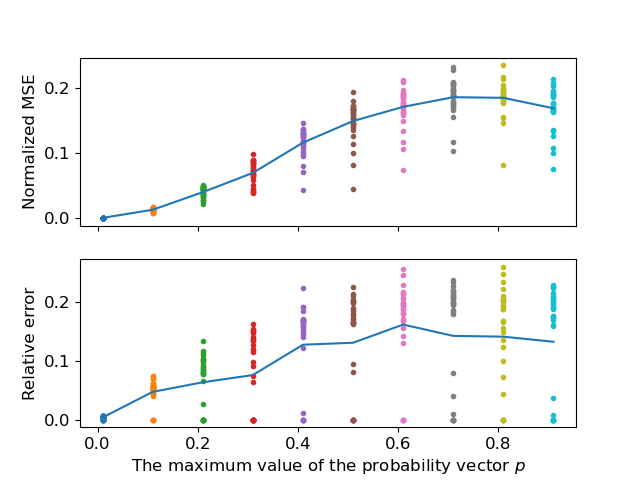}
        \caption{$K=256$: For each term, $1\leq \#flips \leq 8 $ and $-1 \leq \alpha_i\leq 1$. }
    \end{subfigure}
       \caption{\label{Fig:bitflips11}
       Errors defined in Eq.\eqref{Eq:relative error} and \eqref{Eq:nmse error} with different probability choices in $A(p, b)$. }
    \end{figure}

When $\alpha_i$s are no longer positive, any changes in either the control or the system register result in a change in the sum of matrix terms and may drastically change the eigenspectrum of the matrix, especially when we have a large value of  $p_i$. 
For $n=8$ qubits, the change of the relative and the mean squared errors are shown in Figure \ref{Fig:bitflips11} for random matrices of size $2^n$ with $K=2n$ and $K=2^n$ number of summation terms.

\subsubsection{The case where only one of the registers is imperfect}
When only one of the registers is imperfect (i.e., not fault-tolerant), the overall eigenvalue perturbation can be reduced by changing the definition of $A$:
\begin{itemize}
    \item In the case where the control register is imperfect, the number of permutation terms should be reduced by grouping or restructuring the permutation matrices and reducing the control operations, if possible.
    \item In the case where the system register is imperfect but the error rate is small, dividing the summation terms into multiple terms reduces the values of the coefficients. It is known that reducing control register complexity mitigates error propagation, akin to error mitigation in variational algorithms \cite{endo2018practical}. In addition, smaller coefficients $|\alpha_i|$ in these algorithms minimize perturbation impact \cite{cai2023quantum}. Therefore, any bit-flip error cannot cause a substantial change on the overall eigenvalue spectrum.
\end{itemize}

\subsection{Analysis of phase-flips}
\label{Sec:APhaseFlips}
 Phase-flips (modeled as $Z$ gates) invert matrix entries, potentially reversing eigenvalue contributions (e.g., $\alpha_i \rightarrow -\alpha_i$), whereas bit-flips permute entries, preserving non-negativity. Thus, phase-flips induce larger spectral deviations for mixed-sign $\alpha_i$.  
If the rate at which these flips occur is sufficiently small, we can analyze them through the Gershgorin theorem. This at least can show if the circles that cover the eigenvalue change much. And one can conclude if the considered matrix or the system is resilient enough to the errors.

\begin{figure}[t!]
    \centering
    \begin{subfigure}[t]{1\columnwidth}
        \centering
        \includegraphics[width=1\columnwidth]{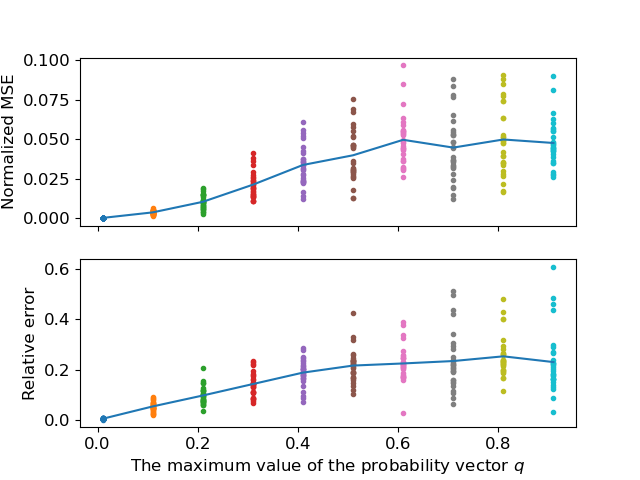}
        \caption{$K=16$: For each term, $1\leq \#flips \leq 8 $ and $-1 \leq \alpha_i\leq 1$.}
    \end{subfigure}
    \begin{subfigure}[t]{1\columnwidth}
        \centering
        \includegraphics[width=1\columnwidth]{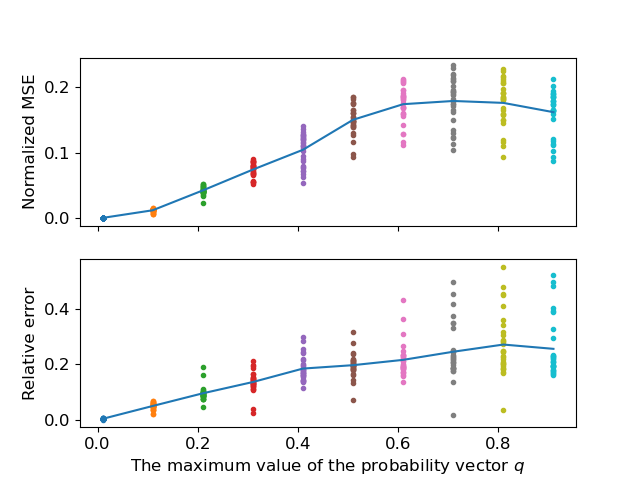}
        \caption{$K=256$: For each term, $1\leq \#flips \leq 8 $ and $-1 \leq \alpha_i\leq 1$. }
    \end{subfigure}
       \caption{\label{Fig:q_bitflips11}
       Errors defined in Eq.\eqref{Eq:relative error} and \eqref{Eq:nmse error} with different probability choices in $A(q, \phi)$. }
\end{figure}

However, when the error rate is high, the matrix changes dramatically. Therefore, any analysis would fail to give meaningful information on the amount of perturbation for any particular eigenvalue.
Figure \ref{Fig:q_bitflips11} shows the errors for random matrices with different choices of the probability $q$ with $\alpha_i\in R$ and $-1 \leq \alpha_i\leq 1$. 
As can be seen from the figure, the results are very similar to the case of bit-flip errors with positive and negative  $\alpha$ coefficients.
Both error types in Eq. \eqref{Eq:Apq} are linear combinations of $I$, $X$, and $Z$ terms. Analogous Gershgorin analysis applies, with phase-flip perturbations bounded by $\Delta R_i \leq 2q_i|\alpha_i|$ (see Appendix \ref{appendixPhaseFlip}).

\section{Analysis on the  fidelity of a quantum state}
\label{Sec:fid}
We will measure the fidelity of an output state generated with $A(p,q, b,\phi)$ by its closeness to the expected output generated with $A$. 
For a random normalized input \ket{\psi}, the original expected output $\ket{\varphi} = {A\ket{\psi}}$, and the erroneous output $\ket{\tilde\varphi} =A(p,q,b,\phi)\ket{\psi}$,  this can be formulated in the normalized form as:
\begin{equation}
\label{Eq:fbraket}
    f_{overlap} = 
    \frac{|\braket{\varphi}{\tilde\varphi}|}{\|\ket{\varphi}\|^2}.
\end{equation}
Note that the normalization is necessary to make this value 1 when \ket{\tilde{\varphi} } = \ket{\varphi}.

We can also measure the fidelity of the output state by using the relative error which is sensitive to the global phase errors. 
For conciseness, if we call the matrix $A(p,q,b, \phi)$ as $B$, we can express the fidelity as:
\begin{equation}
\label{Eq:fre}
\begin{split}
    f_{re} = &\  
   1- \frac{\|\ket{\varphi} - \ket{\tilde\varphi}\|}{\|\ket{\varphi}\|}
    = 1-\frac{\|(A-B)\ket{\psi}\|}{\|A\ket{\psi}\|}. 
\end{split}
\end{equation}
The numerator in the error part measures how close the perturbed matrix to the original matrix. Here note that this problem is commonly known as the matrix nearness problem \cite{higham1988matrix,dhillon2008matrix}.
The relative error term in this equation is very similar to the relative and mean squared eigenvalue errors given in Eq.\eqref{Eq:relative error} and Eq.\eqref{Eq:nmse error}, respectively. 
Therefore they can be considered as indicators for the fidelity. 
For instance, a further analysis of the error term by using the submultiplicative property of the Frobenius or 2-norm yields a very similar equation:

For any matrix \( M \) and vector \( \ket{\psi} \), the operator norm satisfies \cite{golub2013matrix}:
\begin{equation}
\|M \ket{\psi}\| \leq \|M\| \cdot \|\ket{\psi}\|.
\end{equation}
Assuming $\ket{\psi}$ is normalized ($\|\ket{\psi}\| = 1$), using the submultiplicativity of the operator norm, we have:
\begin{equation}
\label{Eqsubmult}
\frac{\|(A - B)\ket{\psi}\|}{\|A\ket{\psi}\|} \leq \frac{\|A - B\| \cdot \|\ket{\psi}\|}{\|A\ket{\psi}\|} = \frac{\|A - B\|}{\|A\ket{\psi}\|}.
\end{equation}
This indicates that the relative error in the output state is bounded by the relative error in the operator, scaled by the inverse of the norm of the transformed state.

We can further relate this to the singular values (or eigenvalues) of the matrices through the properties of the spectral norm. Since the spectral norm satisfies $\|A\ket{\psi}\| \leq \sigma_{max}(A) \|\ket{\psi}\|$ and $\ket{\psi}$ is normalized, we can write:
\begin{equation}
\sigma_{min}(A) \leq \|A\ket{\psi}\| \leq \sigma_{\text{max}}(A),
\end{equation}
where $\sigma_{\text{max}}(\cdot)$ and $\sigma_{\text{min}}(\cdot)$ are the maximum and minimum singular values, respectively.

From these inequalities, we can deduce the following bounds for the relative error:
\begin{equation}
\frac{\|(A - B)\ket{\psi}\|}{\sigma_{max}(A)} \leq
\frac{\|(A - B)\ket{\psi}\|}{\|A\ket{\psi}\|} \leq
\frac{\|(A - B)\|}{\sigma_{min}(A)}.
\end{equation}
A higher value in these terms generally indicates a larger relative error and a lower fidelity.

For the upper bound in the inequality above, we can further write:
\begin{align}
\frac{\|(A - B)\ket{\psi}\|}{\sigma_{min}(A)} \leq &
\frac{\|A - B\|}{\sigma_{min}(A)} =
\frac{\sigma_{max}(A - B)}{\sigma_{min}(A)}\\ & \leq
\frac{\sigma_{max}(A) + \sigma_{max}(B)}{\sigma_{min}(A)}.
\end{align}
This implies that for ill-conditioned \(A\) (that means $\sigma_{min}(A) \ll \sigma_{max}(A)$), even small perturbations \( B \) can amplify relative errors due to the large condition number \(\kappa(A) = \sigma_{max}(A)/\sigma_{min}(A)\).  
This bound is foundational in perturbation analysis for linear systems and matrix computations, where the stability of the system is determined by the condition number of the matrix \cite{golub2013matrix}.

Similarly, for the lower bound of the relative error expressed in terms of \(\sigma_{max}(A)\) in the denominator, we have:
\begin{align}
\frac{\|(A - B)\ket{\psi}\|}{\sigma_{max}(A)} \leq &
\frac{\|A - B\|}{\sigma_{max}(A)} =
\frac{\sigma_{max}(A - B)}{\sigma_{max}(A)}\\ & \leq
\frac{\sigma_{max}(A) + \sigma_{max}(B)}{\sigma_{max}(A)}.
\end{align}

We can also put a lower bound on the relative error using the reverse triangle inequality:
\begin{align}
\|(A - B)\| \geq \left| \|A\| - \|B\| \right| = |\sigma_{max}(A) - \sigma_{max}(B)|.
\end{align}
While this does not directly relate to $\|(A - B)\ket{\psi}\|$, when $\ket{\psi}$ aligns with $A$'s right singular vector corresponding to $\sigma_{max}(A)$ (meaning $\|A\ket{\psi}\| = \sigma_{max}(A)$), it leads to the following lower bound for the relative error:
\begin{align}
\frac{\|(A - B)\ket{\psi}\|}{\|A\ket{\psi}\|}
\geq & \frac{\left| \|A\ket{\psi}\| - \|B\ket{\psi}\| \right|}{\|A\ket{\psi}\|} \\
& = \frac{\left| \sigma_{max}(A) - \|B\ket{\psi}\| \right|}{\sigma_{max}(A)}.
\end{align}
Note that the above analysis aligns with perturbation theory for linear operators, where singular values are used to determine spectral stability \cite{golub2000eigenvalue, bhatia2007perturbation}.

\textbf{Symmetric Matrices:} For symmetric matrices, since $\sigma_{max}(A) = |\lambda_{max}(A)|$, this bound is related to the expression for the relative error given in Eq.\eqref{Eq:relative error}. Furthermore, the analyses in the previous section for eigenvalue perturbation might provide an approximate bound for the relative error which may be observed when the operator is applied to a random input state. 
For example, if \( A \) and \( B \) are symmetric, the eigenvalues of \( A - B \) lie within Gershgorin disks centered at \( (A - B)_{ii} \) with radii \( \Delta R_i = \sum_{j \neq i} |(A - B)_{ij}| \). 
If the Gershgorin disks for \( A - B \) are bounded by \( \Delta R_i \), the spectral norm of \( A - B \) satisfies (\revfix{assuming the disk centers \( (A - B)_{ii} \) are very small or $A$ and $B$ has the same diagonal entries}):
\begin{equation}
\|A - B\| = \max_i |\lambda_i(A - B)| \leq \max_i \Delta R_i.
\end{equation}
For a symmetric matrix \( A \) and normalized \( \ket{\psi} \), the fidelity \( f_{re} \) becomes:
\begin{equation}
f_{re} = 1 - \frac{\|(A - B)\ket{\psi}\|}{\|A\ket{\psi}\|} \geq 1 - \frac{\|A - B\|}{\|A\ket{\psi}\|}.
\end{equation}
If \( \ket{\psi} \) aligns with \( A \)’s dominant eigenvector (\( \|A\ket{\psi}\| = |\lambda_{\text{max}}(A)| \)) and \( \|A - B\| \leq \max_i \Delta R_i \), then:
\begin{equation}
f_{re} \geq 1 - \frac{\max_i \Delta R_i}{|\lambda_{\text{max}}(A)|}.
\end{equation}

We can similarly analyze the case where the fidelity is measured by $f_{overlap}$ given in Eq.\eqref{Eq:fbraket}:
Assuming $A$ and $B$ are real symmetric matrices and $\|\ket{\psi}\| = 1$, we can write the following:
\begin{equation}
\begin{split}
    f_{overlap} = & 
   \frac{\| \braket{\varphi}{\tilde\varphi}\|}{\|\ket{\varphi}\|^2} 
   = \frac{\|\bra{\psi} A^T B \ket{\psi}\|}{\|A\ket{\psi}\|^2}\\
\geq & \frac{\|\bra{\psi} A^T B \ket{\psi}\|}{\|A\|^2}
= \frac{|\lambda_{max}(A^TB)|}{\sigma_{max}(A)^2},  
\end{split}
\end{equation}
where in the last part, we use the Rayleigh-Ritz eigenvalue theorem \cite{golub2013matrix} for the numerator \revfix{by assuming that $A^TB$ has all real eigenvalues and $|\lambda_max(A^TB)|$ represents the eigenvalue with the maximum magnitude} and the definition of the 2-norm for the denominator.

\subsection{Numerical simulation of random states}
In some algorithms the change in the fidelity of the quantum state may be more significant for the overall accuracy of the algorithm.
For instance in the Grover search algorithm\cite{grover1997quantum}, obviously  an error in the marking part of the algorithm leads to marking of a wrong search item. This would then increase the probability of the wrong item which would lead to an unsuccessful run of the algorithm. 
This can be caused by both a bit-flip which would change the marked element and a phase-flip which would mark a wrong element.
However, some algorithms or some parts of them may be more error resilient. For instance, in the Shor's factoring algorithm\cite{shor1994algorithms} all eigenvalues encode the period. Therefore,  a type of error affecting only the period-finding step may not alter the final result if the majority outcome remains valid. Or in the phase estimation algorithm a slight change of the input state may not change the targeted eigenvalue if it does not get close to another eigenstate.

Below, we will observe the fidelity changes by using random input states in the case of bit-flip and phase-flip errors.

\subsection{The fidelities in the case of bit-flips}
Here, we repeat the similar simulations:
 For each error probability $p$, 30 trials are conducted: 
$A(p,b)$ is generated by applying bit-flips to 1–8 random qubits per term in $A$, with $p_i \sim 
\text{Uniform}[0, p_{\text{max}}]$. 
Fidelities $f_{\text{overlap}}$ and $f_{re}$ are averaged over 30 random input states.

\begin{figure}[t!]
    \centering
    \begin{subfigure}[t]{1\columnwidth}
        \centering
        \includegraphics[width=1\columnwidth]{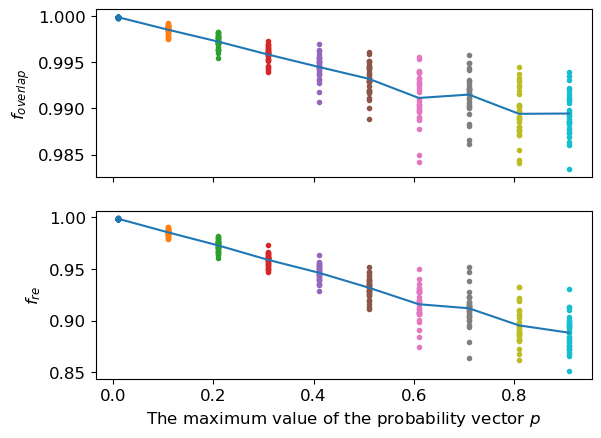}
        \caption{$K=16$: For each term, $1\leq \#flips \leq 8 $ and $0 \leq \alpha_i\leq 1$.}
    \end{subfigure}
    \begin{subfigure}[t]{1\columnwidth}
        \centering
        \includegraphics[width=1\columnwidth]{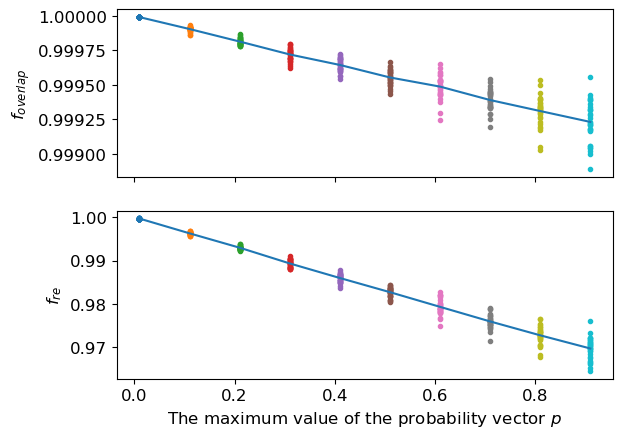}
        \caption{$K=256$: For each term, $1\leq \#flips \leq 8 $ and $0 \leq \alpha_i\leq 1$. }
    \end{subfigure}
       \caption{\label{Fig:fidelity_bitflips01}
      Random state fidelities defined in Eq.\eqref{Eq:fbraket} and \eqref{Eq:fre} with different probability choices in $A(p, b)$. }
\end{figure}

If we choose $\alpha_i$s and the amplitudes of the input state positive, as shown in Fig.\ref{Fig:fidelity_bitflips01} we see the same error resilient observed in Fig.\ref{Fig:bitflips01}.
Therefore, the fidelity of the output state does not diverge from the expected state too  much.

\begin{figure}[t!]
    \centering
    \begin{subfigure}[t]{1\columnwidth}
        \centering
        \includegraphics[width=1\columnwidth]{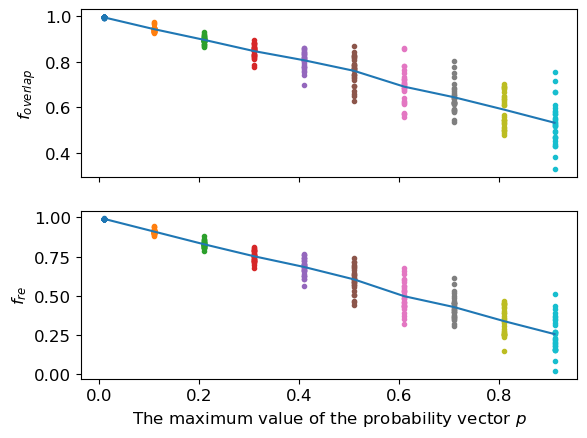}
        \caption{$K=16$: For each term, $1\leq \#flips \leq 8 $ and $-1 \leq \alpha_i\leq 1$.}
    \end{subfigure}
    \begin{subfigure}[t]{1\columnwidth}
        \centering
        \includegraphics[width=1\columnwidth]{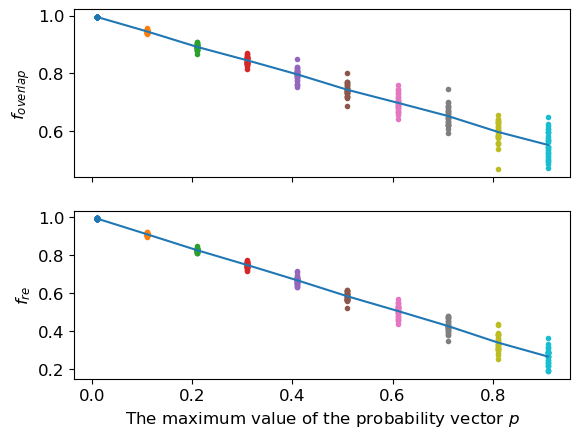}
        \caption{$K=256$: For each term, $1\leq \#flips \leq 8 $ and $-1 \leq \alpha_i\leq 1$. }
    \end{subfigure}
       \caption{\label{Fig:fidelity_bitflips11}
      Random state fidelities defined in Eq.\eqref{Eq:fbraket} and \eqref{Eq:fre} with different probability choices in $A(p, b)$. }
\end{figure}

However, if we choose the amplitudes and $\alpha_i$s in the range [-1,1], the fidelity diverges more from the expected state as drawn in Fig.\ref{Fig:fidelity_bitflips11}.
We see that the amount of the divergence in the case of $f_{re}$ is more than the relative error of the eigenvalue perturbation given in Fig.\ref{Fig:bitflips11} while the behavior is similar. 
Here, note that, as mentioned before, for quantum states $f_{overlap}$ is a more suitable measurement of the fidelity than $f_{re}$ since the latter does not accommodate the global phase differences.

\subsection{The fidelities in the case of phase-flips}
\begin{figure}[t!]
    \centering
    \begin{subfigure}[t]{1\columnwidth}
        \centering
        \includegraphics[width=1\columnwidth]{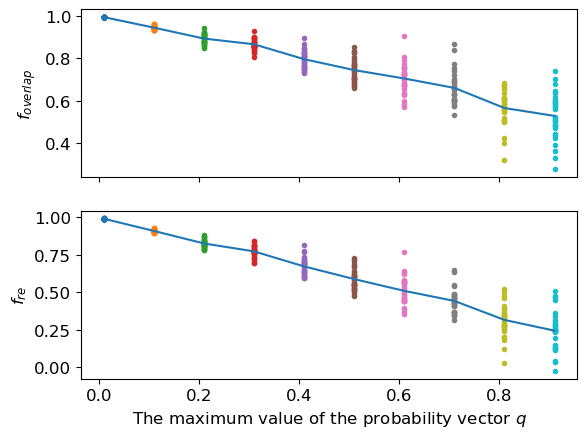}
        \caption{$K=16$: For each term, $1\leq \#flips \leq 8 $ and $-1 \leq \alpha_i\leq 1$.}
    \end{subfigure}
    \begin{subfigure}[t]{1\columnwidth}
        \centering
        \includegraphics[width=1\columnwidth]{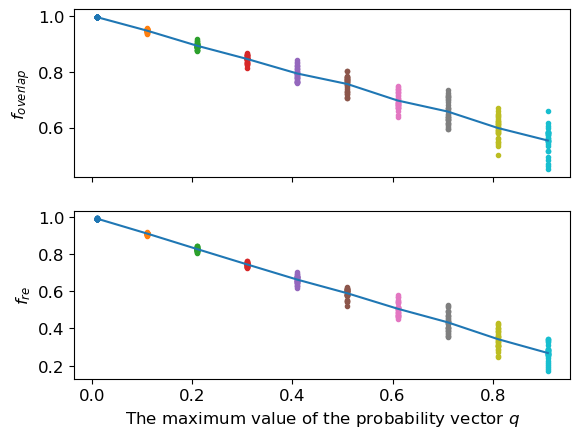}
        \caption{$K=256$: For each term, $1\leq \#flips \leq 8 $ and $-1 \leq \alpha_i\leq 1$. }
    \end{subfigure}
       \caption{\label{Fig:fidelity_phaseflips11}
     Random state fidelities defined in Eq.\eqref{Eq:fbraket} and \eqref{Eq:fre} with different probability choices in $A(q, \phi)$. }
\end{figure}

We repeat the simulations for $A(q,\phi)$. Fig.\ref{Fig:fidelity_phaseflips11} shows the fidelities in this case which is very similar to the case of bit-flips when we choose the amplitudes and $\alpha_i$s in range [-1,1]. 
Fig.\ref{Fig:bitflips11}  and Fig.\ref{Fig:fidelity_bitflips11} exhibit comparable NMSE trends because both error types perturb $A$ by $\mathcal{O}(|\alpha_i|)$ terms. For $|\alpha_i| \leq 1$, phase-flips induce $\pm 2\alpha_i$ deviations, akin to bit-flip permutations.
From this, one may argue that preventing the sign flips  on the matrix elements (this can be caused also by bit-flips) or on the amplitudes are more crucial to get more error resilient algorithms.

\subsection{Sparse input state}
\begin{figure}[t!]
    \centering
    \begin{subfigure}[t]{1\columnwidth}
        \centering
        \includegraphics[width=1\columnwidth]{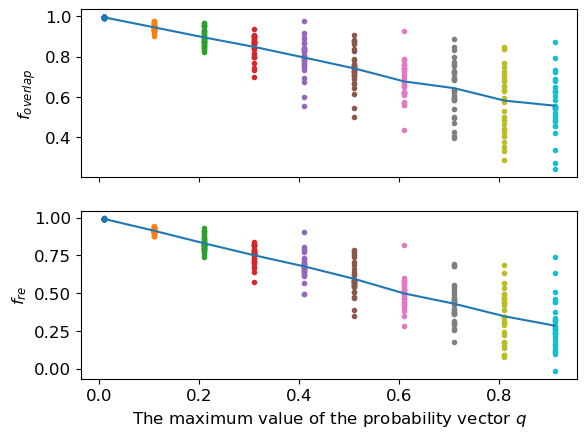}
        \caption{ $A(p, b)$ with $K=16$: For each term, $1\leq \#flips \leq 8 $ and $-1 \leq \alpha_i\leq 1$.}
    \end{subfigure}
    \begin{subfigure}[t]{1\columnwidth}
        \centering
        \includegraphics[width=1\columnwidth]{fig_q_vs_fidelity_A_alpha-11_maxbit8_terms16_sparse0.99.png}
        \caption{$A(q, \phi)$ with $K=16$: For each term, $1\leq \#flips \leq 8 $ and $-1 \leq \alpha_i\leq 1$. }
    \end{subfigure}
       \caption{\label{Fig:fidelity_withsparseinput}
      Fidelities defined in Eq.\eqref{Eq:fbraket} and \eqref{Eq:fre} in the case of sparse random input states. }
\end{figure}

To observe how a sparse input can be affected by the perturbation in matrix $A$, we randomly choose a $99\%$ sparse input state and repeat the last two simulations for bit-flip and phase-flip errors.
Both $A(p,b)$ and $A(q,\phi)$ cases are presented in Fig.\ref{Fig:fidelity_withsparseinput}. As can be seen from the figure, although some cases diverge from the mean value and show better results, the overall picture is similar to the previous simulations. 
Therefore, we can conclude that operator perturbations predominantly govern fidelity loss, though input-sensitive algorithms (e.g., Grover’s search) may exhibit state-dependent error propagation.

\section{Discussion and conclusion}
In this paper, we describe a probabilistic model by writing a matrix as a linear combination of permutations. We have analyzed bit-flip and phase-flip errors and shown that some cases are more error-resilient in terms of eigenvalue perturbation.

The model described in this paper can be used as a basis to study different matrices. As indicated in the introduction, any matrix can be rewritten as a linear combination of permutations. In addition, the coefficients in the matrix can be converted into positive values as follows:
\begin{itemize}
    \item Generalizing permutations: one can push the sign to the permutation. In this case, one needs to implement them in terms of not only $X$ gates but also other gates such as $Z$, $H$, $Y$. 
    \item Converting the matrix into a bistochastic matrix by following Sinkhorn's theorem \cite{sinkhorn1964relationship} and scaling the matrix by diagonal matrices from the left and right. Note that unitary matrices can also be converted by using two unitary matrices \cite{idel2015sinkhorn}.
\end{itemize}

Most quantum numerical algorithms, such as singular value transformation \cite{gilyen2019quantum} or signal processing \cite{low2017optimal}, are modeled through block encoding given in Eq. \eqref{Eq:blockencoding}. Therefore, the error resilience of a matrix can be analyzed before the application of such algorithms.

The model also gives an idea to combine registers with different characteristics. This can also be used to design more error-resilient algorithms (or distributed algorithms) where some local registers are more error-prone than others by shifting parts that impact the eigenvalue spectrum most to the error-resilient registers. For instance, on variational quantum circuits, some terms may be run on non-perfect registers, while others are run on perfect registers.

\section{Data availability}
The simulation code used to generate random permutations and matrices described in Eq.\eqref{Eq:Ap}, \eqref{Eq:Aq}, and \eqref{Eq:Apq} and all figures presented in this paper  is publicly available
on: \url{https://github.com/adaskin/error-analysis-of-sum-of-permutations.git}
\section{Funding}
This project is not funded by any funding agency.
\section{Acknowledgment}
We would like to thank the editor and two anonymous referees for the improvement of the paper and the suggestions for the analyses.
\appendix{
\section{Gershgorin analysis for bit-flip perturbations}
\label{appendixBitFlip}

The bit-flip error modifies each permutation term \( \alpha_i \Pi_i \) in \( A \) as:  
\begin{equation}  
  \Delta A_i = \alpha_i p_i \left( X^{b_i} - I \right) \Pi_i,  
\end{equation}  
where \( X^{b_i} \) applies a Pauli-\(X\) (bit-flip) to qubit(s) \( b_i \). Critically, each permutation matrix \( \Pi_i \) has exactly one ``1'' per row and column. A bit-flip on \( b_i \) swaps entries in \( \Pi_i \), altering at most two elements per row (removing a ``1'' from one position and adding it to another).  
 Suppose \( \Pi_i \) has a 1 at position \( (j, k) \). A bit-flip might move this 1 to a new position \( (j, k') \). This means removing \( \alpha_i \) from column \( k \), and  adding \( \alpha_i \) to column \( k' \).  

The Gershgorin radius \( R_j \) is the sum of absolute off-diagonal elements in row \( j \). Since the perturbation changes two entries, we can write:  
\begin{equation}
    \Delta R_j = |a_{jk} - \alpha_i p_i| + |a_{jk'} + \alpha_i p_i| = 2p_i|\alpha_i|.
\end{equation}  
Here, \( a_{jk} \) and \( a_{jk'} \) are initially zero (off-diagonal elements of \( \Pi_i \)), and the perturbation introduces \( \pm \alpha_i p_i \).

For \( K \) terms in the linear combination \( A = \sum_i \alpha_i \Pi_i \), the total perturbation to \( R_j \) is:  
\begin{equation}  
  \Delta R_j \leq \sum_{i=1}^K 2p_i|\alpha_i|.  
\end{equation} 
The factor of 2 reflects the dual action of bit-flips: removing a contribution from one column and adding it to another. This distinguishes bit-flips from phase-flips (which only alter signs) and ensures the bound aligns with the physical effect of qubit state flips in quantum operations.

\section{Gershgorin analysis for phase-flip perturbations}
\label{appendixPhaseFlip}
The Gershgorin circle theorem states that every eigenvalue of a matrix $A \in \mathbb{C}^{N \times N}$ lies within at least one Gershgorin disc $D_i$, where:  
\begin{equation}
  D_i = \left\{ z \in \mathbb{C} \,:\, |z - a_{ii}| \leq R_i \right\}, \quad R_i = \sum_{j \neq i} |a_{ij}|.  
\end{equation}
For a matrix $A = \sum_{i=1}^K \alpha_i \Pi_i$, phase-flip errors perturb  $A$ to $\tilde{A} = \sum_{i=1}^K \alpha_i \left( (1 - q_i)I + q_i Z^{\phi_i} \right) \Pi_i$. We derive the bound $\Delta R_i \leq 2q_i|\alpha_i|$ for the change in Gershgorin radii as follows:  

The error modifies each term $\alpha_i \Pi_i$  by:  
\begin{equation}
    \Delta A_i = \alpha_i q_i \left( Z^{\phi_i} - I \right) \Pi_i.
\end{equation}
Since \( Z^{\phi_i} \) is a diagonal matrix with entries \( \pm 1 \), the difference \( Z^{\phi_i} - I \) has diagonal entries \( 0 \) or \( -2 \), depending on the qubit \( \phi_i \).

The permutation $\Pi_i$ maps indices $j \to \Pi_i(j)$. For a phase-flip on qubit $\phi_i$, the perturbation $\Delta A_i$ alters the matrix entries in row $j$ by:  
\begin{equation}
|\Delta a_{jk}| \leq 2q_i |\alpha_i| \cdot \delta_{k, \Pi_i(j)},
\end{equation} 
where $\delta_{k, \Pi_i(j)} = 1$ if $k = \Pi_i(j)$, else $0$.

Each term $\Delta A_i$ contributes at most $2q_i|\alpha_i|$ to the Gershgorin radius $R_j$ of row $j$, since only one off-diagonal entry $a_{j,\Pi_i(j)}$ is modified. Summing over all $ K $ terms gives:  
\begin{equation}
\Delta R_j \leq \sum_{i=1}^K 2q_i|\alpha_i|.
\end{equation}
If phase-flips occur independently with $ q_i \ll 1 $, the expected perturbation is:  
\begin{equation}
  \mathbb{E}[\Delta R_j] \leq 2 \sum_{i=1}^K q_i|\alpha_i|.  
\end{equation}
Note that for $q_i \leq q_{\text{max}}$ and 
$|\alpha_i| \leq \alpha_{\text{max}}$, 
this simplifies to 
$\Delta R_j$
$\leq 2K q_{\text{max}}$
$\alpha_{\text{max}}$.  
This  shows that the spectral stability results for bit-flips are equally applicable to phase-flips. This justifies using the same framework to bound eigenvalue deviations under both error types.  

\bibliographystyle{unsrtnat}
\bibliography{paper}

\begin{thebibliography}{32}
\providecommand{\natexlab}[1]{#1}
\providecommand{\url}[1]{\texttt{#1}}
\expandafter\ifx\csname urlstyle\endcsname\relax
  \providecommand{\doi}[1]{doi: #1}\else
  \providecommand{\doi}{doi: \begingroup \urlstyle{rm}\Url}\fi

\bibitem[Kato(2013)]{kato2013perturbation}
Tosio Kato.
\newblock \emph{Perturbation theory for linear operators}, volume 132.
\newblock Springer Science \& Business Media, 2013.

\bibitem[Greenbaum et~al.(2020)Greenbaum, Li, and Overton]{greenbaum2020first}
Anne Greenbaum, Ren-cang Li, and Michael~L Overton.
\newblock First-order perturbation theory for eigenvalues and eigenvectors.
\newblock \emph{SIAM Review}, 62\penalty0 (2):\penalty0 463--482, 2020.

\bibitem[Golub and Van~der Vorst(2000)]{golub2000eigenvalue}
Gene~H Golub and Henk~A Van~der Vorst.
\newblock Eigenvalue computation in the 20th century.
\newblock \emph{Journal of Computational and Applied Mathematics}, 123\penalty0 (1-2):\penalty0 35--65, 2000.

\bibitem[Bhatia(2007)]{bhatia2007perturbation}
Rajendra Bhatia.
\newblock \emph{Perturbation bounds for matrix eigenvalues}.
\newblock SIAM, 2007.

\bibitem[Golub and Van~Loan(2013)]{golub2013matrix}
Gene~H Golub and Charles~F Van~Loan.
\newblock \emph{Matrix computations}.
\newblock JHU press, 2013.

\bibitem[Wilkinson(1960)]{wilkinson1960error}
James~H Wilkinson.
\newblock Error analysis of floating-point computation.
\newblock \emph{Numerische Mathematik}, 2:\penalty0 319--340, 1960.

\bibitem[Veseli{\'c} and Slapni{\v{c}}ar(1993)]{veselic1993floating}
Kre{\v{s}}imar Veseli{\'c} and Ivan Slapni{\v{c}}ar.
\newblock Floating-point perturbations of hermitian matrices.
\newblock \emph{Linear Algebra and its Applications}, 195:\penalty0 81--116, 1993.

\bibitem[Turkeshi and Sierant(2024)]{turkeshi2024error}
Xhek Turkeshi and Piotr Sierant.
\newblock Error-resilience phase transitions in encoding-decoding quantum circuits.
\newblock \emph{Physical Review Letters}, 132\penalty0 (14):\penalty0 140401, 2024.

\bibitem[Huo and Li(2023)]{huo2023error}
Mingxia Huo and Ying Li.
\newblock Error-resilient monte carlo quantum simulation of imaginary time.
\newblock \emph{Quantum}, 7:\penalty0 916, 2023.

\bibitem[Braun et~al.(2022)Braun, Decker, Hegemann, and Kerstan]{braun2022error}
MC~Braun, T~Decker, N~Hegemann, and SF~Kerstan.
\newblock Error resilient quantum amplitude estimation from parallel quantum phase estimation.
\newblock \emph{arXiv preprint arXiv:2204.01337}, 2022.

\bibitem[Xiong et~al.(2021)Xiong, Ng, and Hanzo]{xiong2021quantum}
Yifeng Xiong, Soon~Xin Ng, and Lajos Hanzo.
\newblock Quantum error mitigation relying on permutation filtering.
\newblock \emph{IEEE Transactions on Communications}, 70\penalty0 (3):\penalty0 1927--1942, 2021.

\bibitem[Lowe et~al.(2021)Lowe, Gordon, Czarnik, Arrasmith, Coles, and Cincio]{lowe2021unified}
Angus Lowe, Max~Hunter Gordon, Piotr Czarnik, Andrew Arrasmith, Patrick~J Coles, and Lukasz Cincio.
\newblock Unified approach to data-driven quantum error mitigation.
\newblock \emph{Physical Review Research}, 3\penalty0 (3):\penalty0 033098, 2021.

\bibitem[Cai et~al.(2023)Cai, Babbush, Benjamin, Endo, Huggins, Li, McClean, and O’Brien]{cai2023quantum}
Zhenyu Cai, Ryan Babbush, Simon~C Benjamin, Suguru Endo, William~J Huggins, Ying Li, Jarrod~R McClean, and Thomas~E O’Brien.
\newblock Quantum error mitigation.
\newblock \emph{Reviews of Modern Physics}, 95\penalty0 (4):\penalty0 045005, 2023.

\bibitem[Takagi et~al.(2022)Takagi, Endo, Minagawa, and Gu]{takagi2022fundamental}
Ryuji Takagi, Suguru Endo, Shintaro Minagawa, and Mile Gu.
\newblock Fundamental limits of quantum error mitigation.
\newblock \emph{npj Quantum Information}, 8\penalty0 (1):\penalty0 114, 2022.

\bibitem[Gily{\'e}n et~al.(2019)Gily{\'e}n, Su, Low, and Wiebe]{gilyen2019quantum}
Andr{\'a}s Gily{\'e}n, Yuan Su, Guang~Hao Low, and Nathan Wiebe.
\newblock Quantum singular value transformation and beyond: exponential improvements for quantum matrix arithmetics.
\newblock In \emph{Proceedings of the 51st Annual ACM SIGACT Symposium on Theory of Computing}, pages 193--204, 2019.

\bibitem[Birkhoff(1946)]{birkhoff1946tres}
Garrett Birkhoff.
\newblock Tres observaciones sobre el algebra lineal.
\newblock \emph{Univ. Nac. Tucuman, Ser. A}, 5:\penalty0 147--154, 1946.

\bibitem[Sinkhorn(1964)]{sinkhorn1964relationship}
Richard Sinkhorn.
\newblock A relationship between arbitrary positive matrices and doubly stochastic matrices.
\newblock \emph{The Annals of Mathematical Statistics}, 35\penalty0 (2):\penalty0 876--879, 1964.

\bibitem[Kulkarni et~al.(2017)Kulkarni, Lee, and Singh]{kulkarni2017minimum}
Janardhan Kulkarni, Euiwoong Lee, and Mohit Singh.
\newblock Minimum birkhoff-von neumann decomposition.
\newblock In \emph{International Conference on Integer Programming and Combinatorial Optimization}, pages 343--354. Springer, 2017.

\bibitem[Daskin(2024)]{daskin2024quantum}
Ammar Daskin.
\newblock A quantum compiler design method by using linear combinations of permutations.
\newblock \emph{arXiv preprint arXiv:2404.18226}, 2024.

\bibitem[Idel and Wolf(2015)]{idel2015sinkhorn}
Martin Idel and Michael~M Wolf.
\newblock Sinkhorn normal form for unitary matrices.
\newblock \emph{Linear Algebra and its Applications}, 471:\penalty0 76--84, 2015.

\bibitem[Daskin and Kais(2018)]{daskin2018generalized}
Ammar Daskin and Sabre Kais.
\newblock A generalized circuit for the hamiltonian dynamics through the truncated series.
\newblock \emph{Quantum Information Processing}, 17:\penalty0 1--19, 2018.

\bibitem[Kitaev(1997)]{kitaev1997quantum}
A~Yu Kitaev.
\newblock Quantum computations: algorithms and error correction.
\newblock \emph{Russian Mathematical Surveys}, 52\penalty0 (6):\penalty0 1191, 1997.

\bibitem[Devitt et~al.(2013)Devitt, Munro, and Nemoto]{devitt2013quantum}
Simon~J Devitt, William~J Munro, and Kae Nemoto.
\newblock Quantum error correction for beginners.
\newblock \emph{Reports on Progress in Physics}, 76\penalty0 (7):\penalty0 076001, 2013.

\bibitem[Nielsen and Chuang(2010)]{nielsen2010quantum}
Michael~A Nielsen and Isaac~L Chuang.
\newblock \emph{Quantum omputation and quantum information}.
\newblock Cambridge University Press, 2010.

\bibitem[Brualdi(1988)]{brualdi1988some}
Richard~A Brualdi.
\newblock Some applications of doubly stochastic matrices.
\newblock \emph{Linear Algebra and its Applications}, 107:\penalty0 77--100, 1988.

\bibitem[Horn and Johnson(2012)]{horn2012matrix}
Roger~A Horn and Charles~R Johnson.
\newblock \emph{Matrix analysis}.
\newblock Cambridge University Press, 2012.

\bibitem[Endo et~al.(2018)Endo, Benjamin, and Li]{endo2018practical}
Suguru Endo, Simon~C Benjamin, and Ying Li.
\newblock Practical quantum error mitigation for near-future applications.
\newblock \emph{Physical Review X}, 8\penalty0 (3):\penalty0 031027, 2018.

\bibitem[Higham(1988)]{higham1988matrix}
Nicholas~J Higham.
\newblock \emph{Matrix nearness problems and applications}.
\newblock University of Manchester. Department of Mathematics, 1988.

\bibitem[Dhillon and Tropp(2008)]{dhillon2008matrix}
Inderjit~S Dhillon and Joel~A Tropp.
\newblock Matrix nearness problems with bregman divergences.
\newblock \emph{SIAM Journal on Matrix Analysis and Applications}, 29\penalty0 (4):\penalty0 1120--1146, 2008.

\bibitem[Grover(1997)]{grover1997quantum}
Lov~K Grover.
\newblock Quantum computers can search arbitrarily large databases by a single query.
\newblock \emph{Physical Review Letters}, 79\penalty0 (23):\penalty0 4709, 1997.

\bibitem[Shor(1994)]{shor1994algorithms}
Peter~W Shor.
\newblock Algorithms for quantum computation: discrete logarithms and factoring.
\newblock In \emph{Proceedings 35th Annual Symposium on Foundations of Computer Science}, pages 124--134. Ieee, 1994.

\bibitem[Low and Chuang(2017)]{low2017optimal}
Guang~Hao Low and Isaac~L Chuang.
\newblock Optimal hamiltonian simulation by quantum signal processing.
\newblock \emph{Physical Review Letters}, 118\penalty0 (1):\penalty0 010501, 2017.

\end{thebibliography}

\end{document}